\documentclass[11pt, a4paper]{article}

\ifcsname dgB@loaded\endcsname\expandafter\endinput\fi
\expandafter\let\csname dgB@loaded\endcsname\empty

\usepackage{ifthen}
\usepackage{etoolbox}%

\makeatletter
\@ifundefined{ifdelim_grow}{%
\newboolean{delim_grow}\setboolean{delim_grow}{true}%
}{}
\@ifundefined{ifare_slides}{%
\newboolean{are_slides}%
\setboolean{are_slides}{false}%
}{}
\@ifundefined{ifdg_impose_style}{%
\newboolean{dg_impose_style}\setboolean{dg_impose_style}{true}%
}{}
\@ifundefined{ifdg_load_tikz}{%
\newboolean{dg_load_tikz}\setboolean{dg_load_tikz}{true}%
}{}
\makeatother

\usepackage{amssymb}
\usepackage{amsmath}%

\usepackage{amsthm}

\usepackage{latexsym}

\ifthenelse{\boolean{are_slides}}{}{\usepackage{enumitem}}

\usepackage{units}%

\usepackage{filemod}%

\usepackage[nolabel]{fnbreak}%

\usepackage{psfrag}%

\makeatletter
\@ifpackageloaded{xcolor}{%
\@ifundefinedcolor{BrickRed}{\input{dvipsnam.def}}{}%
}{\usepackage[dvipsnames]{xcolor}}
\makeatother
\ifthenelse{\boolean{dg_load_tikz}}{\usepackage{tikz}}{}

\makeatletter
\@ifpackageloaded{url}{}{\PassOptionsToPackage{hyphens,spaces,obeyspaces}{url}}
\makeatother
\usepackage{url}%

\usepackage{graphicx}

\usepackage{eso-pic}%

\usepackage{marginnote}%

\usepackage{caption}%

\usepackage{calc}%

\usepackage{soul}%
\setul{1.25pt}{.16pt}%

\usepackage{bm}%

\usepackage{newunicodechar}%
\newunicodechar{⇒}{\ensuremath{\Rightarrow}}
\newunicodechar{ꞏ}{\scalebox{0.75}{\ensuremath{\bullet}}}

\ifthenelse{\boolean{dg_impose_style}}{%
\usepackage{tempora}%

\usepackage[T1,OT2,T2A]{fontenc}

\usepackage[utf8]{inputenc}

\usepackage[shorthands=off,greek,czech,ukrainian,english]{babel}

\DeclareTextSymbolDefault{\CYRYAT}{OT2}
\DeclareTextSymbolDefault{\cyryat}{OT2}
\DeclareTextSymbolDefault{\CYRFITA}{OT2}
\DeclareTextSymbolDefault{\cyrfita}{OT2}
\DeclareTextSymbolDefault{\CYRIZH}{OT2}
\DeclareTextSymbolDefault{\cyrizh}{OT2}

\fontencoding{OT1}\selectfont%
\usepackage[onlymath]{sansmathfonts}
\fontencoding{T2A}\selectfont

}{}%

\usepackage{microtype}%

\usepackage{mathrsfs}%

\usepackage{calligra}%

\usepackage{pdfrender}%

\usepackage{hyperref}%
\setpdflinkmargin{0.75pt}%
\ifthenelse{\boolean{dg_impose_style}}{%
\hypersetup{
colorlinks,
linkcolor={red!55!black},
citecolor={blue!55!black},
urlcolor={blue!35!black}
}
}{}

\ifthenelse{\boolean{are_slides}}{
\usepackage{beamerthemesplit}

\usepackage{ragged2e}%

\usetheme[secheader]{Boadilla}

\hypersetup{hidelinks}%

\usenavigationsymbolstemplate{}%
}{}%

\ifthenelse{\boolean{are_slides}}{
\newcommand{\MyChangeMathMargins}{%
\setlength{\abovedisplayskip}{\abovedisplayshortskip + 3pt}%
\setlength{\belowdisplayskip}{\abovedisplayshortskip + 4pt}%
}

\setul{1.5pt}{0.65pt}%
}{
\newcommand{\MyChangeMathMargins}{%
\setlength{\abovedisplayskip}{\abovedisplayshortskip + 4pt}%
\setlength{\belowdisplayskip}{\abovedisplayshortskip + 5pt}%
}

\ifthenelse{\boolean{dg_impose_style}}{%
\setlength\textheight{9.5in}

\setlength\topmargin{\dimexpr(\paperheight-\textheight
-\headheight-\headsep-\footskip)/2 - 1in\relax}

\setlength\textwidth{6.2in}

\setlength\oddsidemargin{\dimexpr(\paperwidth-\textwidth)/2 - 1in\relax}
\setlength\evensidemargin{\oddsidemargin}

\pdfpageheight=\paperheight
\pdfpagewidth=\paperwidth
}{}%
}%

\makeatletter

\newcommand{\NewcommandThreeArgsTwoOpt}[6][*]{%
\ifthenelse{\equal{}{#1}}%
{\let\comm@TMP\DeclareDocumentCommand}%
{\let\comm@TMP\NewDocumentCommand}%
\expandafter\def\csname\expandafter\@gobble\string#2@nopresq@nopostsq\endcsname##1##2{#3}%
\expandafter\def\csname\expandafter\@gobble\string#2@presq@nopostsq\endcsname##1##2{#4}%
\expandafter\def\csname\expandafter\@gobble\string#2@nopresq@postsq\endcsname##1##2##3{#5}%
\expandafter\def\csname\expandafter\@gobble\string#2@presq@postsq\endcsname##1##2##3{#6}%
\comm@TMP{#2}{o m o}{%
\IfNoValueTF{##1}{%
\IfNoValueTF{##3}%
{\csname\expandafter\@gobble\string#2@nopresq@nopostsq\endcsname{}{##2}}%
{\csname\expandafter\@gobble\string#2@nopresq@postsq\endcsname{}{##2}{##3}}%
}{%
\IfNoValueTF{##3}%
{\csname\expandafter\@gobble\string#2@presq@nopostsq\endcsname{##1}{##2}}%
{\csname\expandafter\@gobble\string#2@presq@postsq\endcsname{##1}{##2}{##3}}%
}%
}%
}

\newcommand{\NewcommandTwoArgsLastOpt}[4][*]{%
\ifthenelse{\equal{}{#1}}%
{\let\comm@TMP\DeclareDocumentCommand}%
{\let\comm@TMP\NewDocumentCommand}%
\expandafter\def\csname\expandafter\@gobble\string#2@nopostsq\endcsname##1{#3}%
\expandafter\def\csname\expandafter\@gobble\string#2@postsq\endcsname##1##2{#4}%
\comm@TMP{#2}{m o}{%
\IfNoValueTF{##2}%
{\csname\expandafter\@gobble\string#2@nopostsq\endcsname{##1}}%
{\csname\expandafter\@gobble\string#2@postsq\endcsname{##1}{##2}}%
}%
}

\newcommand{\NewcommandFourArgsTwoOpt}[6][*]{%
\ifthenelse{\equal{}{#1}}%
{\let\comm@TMP\DeclareDocumentCommand}%
{\let\comm@TMP\NewDocumentCommand}%
\expandafter\def\csname\expandafter\@gobble\string#2@nopresq@nopostsq\endcsname##1##2##3{#3}%
\expandafter\def\csname\expandafter\@gobble\string#2@presq@nopostsq\endcsname##1##2##3{#4}%
\expandafter\def\csname\expandafter\@gobble\string#2@nopresq@postsq\endcsname##1##2##3##4{#5}%
\expandafter\def\csname\expandafter\@gobble\string#2@presq@postsq\endcsname##1##2##3##4{#6}%
\comm@TMP{#2}{o m m o}{%
\IfNoValueTF{##1}{%
\IfNoValueTF{##4}%
{\csname\expandafter\@gobble\string#2@nopresq@nopostsq\endcsname{}{##2}{##3}}%
{\csname\expandafter\@gobble\string#2@nopresq@postsq\endcsname{}{##2}{##3}{##4}}%
}{%
\IfNoValueTF{##4}%
{\csname\expandafter\@gobble\string#2@presq@nopostsq\endcsname{##1}{##2}{##3}}%
{\csname\expandafter\@gobble\string#2@presq@postsq\endcsname{##1}{##2}{##3}{##4}}%
}%
}%
}

\newcommand{\NewcommandFourArgsTwoOptDefa}[5][*]{%
\ifthenelse{\equal{}{#1}}%
{\let\comm@TMP\DeclareDocumentCommand}%
{\let\comm@TMP\NewDocumentCommand}%
\comm@TMP{#2}{O{#3} m m O{#4}}{#5}%
}

\newcommand{\ifndef}[2]{\@ifundefined{#1}{#2}{}}

\newcommand{\IfMathMode}[2]{\ifmmode{#1}\else{#2}\fi}

\DeclareRobustCommand\bk{\@ifnextchar[%
{\lf\langle \bk@bra{}}%
{\@ifnextchar<%
{\lf.\bk@op}
{\lf| \bk@ket{}}%
}%
}

\def\bk@bra#1[#2]{\def\bk@tmp{#2}\ifx\bk@tmp\empty{#1}\else{#2}\fi\@ifnextchar[%
{\md|\hspace{-1.5pt}\md\langle \bk@bra{\def\bk@tmp{#2}\ifx\bk@tmp\empty{#1}\else{#2}\fi}}%
{\@ifnextchar\bgroup%
{\md| \bk@ket{\def\bk@tmp{#2}\ifx\bk@tmp\empty{#1}\else{#2}\fi}}%
{\@ifnextchar<%
{\md| \bk@op}%
{\rt|}%
}%
}%
}

\def\bk@ket#1#2{\def\bk@tmp{#2}\ifx\bk@tmp\empty{#1}\else{#2}\fi\@ifnextchar[%
{\md\rangle\hspace{-1.5pt}\md\langle \bk@bra{\def\bk@tmp{#2}\ifx\bk@tmp\empty{#1}\else{#2}\fi}}%
{\@ifnextchar\bgroup%
{\md\rangle\hspace{-1.5pt}\md| \bk@ket{\def\bk@tmp{#2}\ifx\bk@tmp\empty{#1}\else{#2}\fi}}%
{\@ifnextchar<%
{\md\rangle \bk@op}%
{\rt\rangle}%
}%
}%
}

\def\bk@op<#1>{#1 \@ifnextchar[%
{\md\langle \bk@bra{}}%
{\@ifnextchar\bgroup%
{\md| \bk@ket{}}%
{\@ifnextchar<%
{\tm \bk@op}%
{\rt.}%
}%
}%
}

\newcommand\Cases[1][0pt]{%
\def\Case@skipam{#1}%
\left\{\!\!\!\begin{array}{ll}\Cases@continue%
}

\def\Cases@continue#1#2{\@ifnextchar\bgroup%
{#1 &\txt{#2}\\[\Case@skipam] \Cases@continue}%
{#1 &\txt{#2}\end{array}\Cases@end}%
}

\def\Cases@end{\@ifnextchar[%
{\Cases@end@postsq}%
{\right.}
}

\def\Cases@end@postsq[#1]{\ifthenelse{\equal{\}}{#1}}
{\!\!\right\}}%
{\right.}
}

\newcommand{\MyMakeTheoMacros}[3]{
\expandafter\newcommand\csname\expandafter\@gobble\string#2NostarNoname@DGaux\endcsname[2][]
{\ifthenelse{\equal{}{##1}}%
{\begin{#1}~##2 \end{#1}}%
{\begin{#1}\label{##1}~##2\end{#1}}%
}
\expandafter\newcommand\csname\expandafter\@gobble\string#2StarNoname@DGaux\endcsname[1]
{\begin{#1*}~##1 \end{#1*}}
\newcommand#2{\expandafter\@ifstar%
\expandafter{\csname\expandafter\@gobble\string#2StarNoname@DGaux\endcsname}%
{\csname\expandafter\@gobble\string#2NostarNoname@DGaux\endcsname}%
}

\expandafter\newcommand\csname\expandafter\@gobble\string#2NostarName@DGaux\endcsname[3][]
{\ifthenelse{\equal{}{##1}}%
{\begin{#1}[\e{##2}]~##3 \end{#1}}%
{\begin{#1}[\e{##2}]\label{##1}~##3\end{#1}}%
}
\expandafter\newcommand\csname\expandafter\@gobble\string#2StarName@DGaux\endcsname[2]
{\begin{#1*}[\e{##1}]~##2 \end{#1*}}
\newcommand#3{\expandafter\@ifstar%
\expandafter{\csname\expandafter\@gobble\string#2StarName@DGaux\endcsname}
{\csname\expandafter\@gobble\string#2NostarName@DGaux\endcsname}%
}
}

\DeclareRobustCommand\dgref[2]{\@ifnextchar[%
{#2\dgref@presq}%
{#1\dgref@nopresq}%
}

\def\dgref@presq[#1]{\@ifnextchar[%
{\ref{#1}, \dgref@presq}
{\ref{#1} and~\dgref@nopresq}
}

\def\dgref@nopresq#1{\ref{#1}}

\DeclareRobustCommand\bref{\@ifnextchar[%
{\bref@presq}%
{\bref@nopresq}%
}

\def\bref@presq[#1]{\@ifnextchar[%
{(\ref{#1}), \bref@presq}
{(\ref{#1}) and~\bref@nopresq}
}

\def\bref@nopresq#1{(\ref{#1})}

\makeatother

\newcommand{\NewOp}[2]{
\newcommand{#1}[2][]{\ensuremath{\operatorname{#2}##1\ifthenelse{\equal{}{##2}}{}{\lf(##2\rt)}}}
}

\newcommand{\NewId}[2]{
\newcommand{#1}[1][]{\ensuremath{\ifthenelse{\equal{}{##1}}{\operatorname{\mathit{#2}}}{\operatorname{\mathit{#2##1}}}}}
}

\newcommand{\NewIdR}[2]{
\newcommand{#1}[1][]{\ensuremath{\ifthenelse{\equal{}{##1}}{\operatorname{\mathrm{#2}}}{\operatorname{\mathrm{#2##1}}}}}
}

\newcommand{\NewIdSm}[2]{\NewId{#1}%
{\mathchoice{#2}{#2}{\scriptscriptstyle{#2}}{\scalebox{0.75}{$\scriptscriptstyle{#2}$}}}%
}

\NewIdSm{\tp}{\top}
\NewIdSm{\bt}{\bot}

\newcommand{\NewM}[2]{
\newcommand{#1}{\ensuremath{#2}}
}

\NewM{\OI}{\set{0,1}}
\NewM{\II}{\set{-1,1}}
\NewM{\TF}{\set{\tp,\bt}}

\newcommand{\MyMakeRefMacros}[3]{\newcommand{#1}{\dgref{#2}{#3}}}

\ifndef{theorem}{}
\ifndef{theorem*}{\newtheorem*{theorem*}{Theorem}}

\MyMakeTheoMacros{theorem}{\theo}{\ntheo}

\MyMakeRefMacros{\theoref}{Theorem~}{Theorems~}

\ifndef{definition}{}
\ifndef{definition*}{\newtheorem*{definition*}{Definition}}

\MyMakeTheoMacros{definition}{\defi}{\ndefi}

\MyMakeRefMacros{\defiref}{Definition~}{Definitions~}

\ifndef{claim}{}
\ifndef{claim*}{\newtheorem*{claim*}{Claim}}

\MyMakeTheoMacros{claim}{\clm}{\nclm}

\MyMakeRefMacros{\clmref}{Claim~}{Claims~}

\ifndef{lemma}{}
\ifndef{lemma*}{\newtheorem*{lemma*}{Lemma}}

\MyMakeTheoMacros{lemma}{\lem}{\nlem}

\MyMakeRefMacros{\lemref}{Lemma~}{Lemmas~}

\ifndef{corollary}{}
\ifndef{corollary*}{\newtheorem*{corollary*}{Corollary}}

\MyMakeTheoMacros{corollary}{\crl}{\ncrl}

\MyMakeRefMacros{\crlref}{Corollary~}{Corollaries~}

\ifndef{fact}{}
\ifndef{fact*}{\newtheorem*{fact*}{Fact}}

\MyMakeTheoMacros{fact}{\fct}{\nfct}

\MyMakeRefMacros{\fctref}{Fact~}{Facts~}

\ifndef{question}{}
\ifndef{question*}{\newtheorem*{question*}{Question}}

\MyMakeTheoMacros{question}{\quest}{\nquest}

\MyMakeRefMacros{\questref}{Question~}{Questions~}

\ifndef{construction}{}
\ifndef{construction*}{\newtheorem*{construction*}{Construction}}

\MyMakeTheoMacros{construction}{\constr}{\nconstr}

\MyMakeRefMacros{\constrref}{Construction~}{Constructions~}

\renewcommand{\qedsymbol}{$\blacksquare$}

\DeclareRobustCommand{\prflabel}[1]{%
\ifstrempty{#1}%
{\renewcommand{\qedsymbol}{$\blacksquare$}}%
{\renewcommand{\qedsymbol}{$\blacksquare_{\mbox{\it{\scriptsize{#1}}}}$}}%
}
\newcommand{\prfstart}[1][]{\ifthenelse{\equal{}{#1}}%
{\begin{proof}[\underline{\proofname}]\prflabel{}}%
{\begin{proof}[\underline{\proofname\ #1}]\prflabel{#1}}%
}

\newcommand{\prfend}[1][*]{%
\ifthenelse{\equal{*}{#1}}{}{\prflabel{#1}}%
\end{proof}%
}

\newcommand{\abstart}{\begin{abstract}}
\newcommand{\abend}{\end{abstract}}

\newcommand{\NewSectLike}[3]{
\NewcommandThreeArgsTwoOpt{#1}
{\ifthenelse{\equal{*}{##2}}{#2*}{#2{##2}}}
{#2{##2}\label{##1}}
{#2[##3]{##2}#3{##3}}
{#2[##3]{##2}\label{##1}#3{##3}}
}

\NewSectLike{\sect}{\section}{\sectionmark}

\NewSectLike{\ssect}{\subsection}{\subsectionmark}

\NewSectLike{\sssect}{\subsubsection}{\subsubsectionmark}

\MyMakeRefMacros{\sref}{Section~}{Sections~}
\MyMakeRefMacros{\ssref}{Subsection~}{Subsections~}

\newcommand{\toc}[3][]{%
\ifthenelse{\equal{chapter}{#2}}{\newpage}{}%
\ifthenelse{\equal{part}{#2}}{\newpage}{}%
\ifthenelse{\equal{}{#1}}%
{\phantomsection\addcontentsline{toc}{#2}{#3}}%
{\phantomsection\refstepcounter{#2}\label{#1}\addcontentsline{toc}{#2}{#3}}%
}

\newcommand{\toct}[1]{\toc{section}{#1}}

\makeatletter
\newif\ifdg@bibtoc
\newcommand{\bibtoc}{%
\ifdg@bibtoc\else
\apptocmd{\thebibliography}{\toct{\refname}}%
{\dg@bibtoctrue}%
{\PackageError{dgB}{Cannot add the bibliography TOC entry}%
{Call \string\bibtoc\space after bibliography-related packages.}}%
\fi
}
\makeatother

\newcommand{\bibstart}[1][99]{\bibtoc\begin{thebibliography}{#1}}
\newcommand{\bibend}{\end{thebibliography}}

\MyMakeRefMacros{\fnref}{Footnote~}{Footnotes~}

\newcommand{\m}[2][]{\MyChangeMathMargins%
\ifthenelse{\boolean{delim_grow}}{\delimiterfactor=1001}{}%
\ifthenelse{\equal{}{#1}}%
{\begin{align*} #2 \end{align*}}%
{\ifthenelse{\equal{P}{#1}}%
{\allowdisplaybreaks\begin{align*} #2%
\end{align*}\interdisplaylinepenalty=10000}%
{\ifthenelse{\equal{I}{#1}}%
{$#2$}%
{\begin{align}\begin{split}\label{#1} #2 \end{split}\end{align}}%
}%
}%
}

\makeatletter

\let\dgampersand\&

\DeclareRobustCommand\&{%
\new@ifnextchar[%
{\dgsep@reposit}%
{\dgampersand}%
}

\def\dgsep@reposit[#1]{\hspace{#1}&\hspace{-#1}}

\makeatother

\mathchardef\breakingcomma\mathcode`\,
{\catcode`,=\active
\gdef,{\breakingcomma\discretionary{}{}{}}
}

\newcommand{\lf}{\mathopen{}\mathclose\bgroup\left}
\newcommand{\rt}{\aftergroup\egroup\right}

\providecommand{\middle}{\big}
\newcommand{\md}{\middle}

\newcommand{\ud}{\vphantom{|_1^1}}

\makeatletter

\newcommand{\NewPar}[4][*]{
\ifthenelse{\equal{}{#1}}
{\def\dg@parfactory{\NewcommandThreeArgsTwoOpt[]}}
{\def\dg@parfactory{\NewcommandThreeArgsTwoOpt}}
\dg@parfactory{#2}
{\lf#3{##2}\rt#4}
{#3{##2}#4}
{\lf#3{##2}\rt#4_{\!##3}}
{#3{##2}#4_{\!##3}}
}
\makeatother

\NewPar{\lrp}{(}{)}
\NewPar{\lra}{\langle}{\rangle}
\NewPar[]{\sz}{|}{|}
\NewPar[]{\norm}{\|}{\|}
\NewPar[]{\ceil}{\lceil}{\rceil}
\NewPar[]{\floor}{\lfloor}{\rfloor}
\NewPar{\round}{\lfloor}{\rceil}

\NewcommandThreeArgsTwoOpt[]{\set}
{\lf\{#2\rt\}}
{}
{\lf\{#2\md|\ud#3\rt\}}
{\lf\{#2\ud#1\ud#3\rt\}}

\definecolor{Red}{rgb}{1,0,0}
\definecolor{DarkRed}{rgb}{0.54,0,0}
\definecolor{DarkGreen}{rgb}{0, 0.52, 0.05}
\definecolor{LightBlue}{rgb}{0.122,0.016,0.855}

\definecolor{dgProbar}{rgb}{0.3,0.3,0.3}%
\definecolor{dgPreBack}{RGB}{209,211,214}%

\DeclareTextFontCommand{\bemph}{\bfseries}
\DeclareTextFontCommand{\ibemph}{\bfseries\em}
\DeclareTextFontCommand{\ttl}{\ttfamily}

\ifthenelse{\boolean{are_slides}}{

\newcommand{\e}{\ei}
\newcommand<>{\ei}[1]{\IfMathMode%
{\only#2{\textcolor{BrickRed}}{#1}}%
{\emph{\only#2{\textcolor{BrickRed}}{#1}}}%
}
\newcommand<>{\eb}[1]{\IfMathMode%
{\only#2{\textcolor{BrickRed}}{\pmb{#1}}}%
{\bemph{\only#2{\textcolor{BrickRed}}{#1}}}%
}
\newcommand<>{\ebi}[1]{\IfMathMode%
{\only#2{\textcolor{BrickRed}}{\pmb{#1}}}%
{\ibemph{\only#2{\textcolor{BrickRed}}{#1}}}%
}
\newcommand<>{\eu}[1]{\IfMathMode%
{\only#2{\textcolor{BrickRed}}{\txt{\ul{#1}}}}%
{\only#2{\textcolor{BrickRed}}{\ul{#1}}}%
}
\newcommand<>{\eui}[1]{\IfMathMode%
{\only#2{\textcolor{BrickRed}}{\txt{\ul{#1}}}}%
{\emph{\only#2{\textcolor{BrickRed}}{\ul{#1}}}}%
}
\newcommand<>{\ebu}[1]{\IfMathMode%
{\only#2{\textcolor{BrickRed}}{\underline{\pmb{#1}}}}%
{\bemph{\only#2{\textcolor{BrickRed}}{\ul{#1}}}}%
}
\newcommand<>{\ebui}[1]{\IfMathMode%
{\only#2{\textcolor{BrickRed}}{\underline{#1}}}%
{\ibemph{\only#2{\textcolor{BrickRed}}{\ul{#1}}}}%
}
}{%
\newcommand{\e}{\emph}
\newcommand{\eb}{\bemph}
\newcommand{\ebi}{\ibemph}
\newcommand{\eu}[1]{\ttl{\ul{#1}}}
\newcommand{\eui}[1]{\emph{\ul{#1}}}
\newcommand{\ebu}[1]{\bemph{\ul{#1}}}
\newcommand{\ebui}[1]{\ibemph{\ul{#1}}}
}%

\newcommand{\txt}[1]{\textrm{#1}}%

\newcommand{\tm}{\cdot}

\NewcommandFourArgsTwoOptDefa{\overlay}
{0mu}{1}
{
\begingroup
\mathchoice{%
\bgroup
\scalebox{#4}{\ooalign{$\displaystyle#2\mkern#1$\cr
\hidewidth{$\displaystyle\mkern#1#3$}\hidewidth}}
\egroup
}{%
\bgroup
\scalebox{#4}{\ooalign{$\textstyle#2\mkern#1$\cr
\hidewidth{$\textstyle\mkern#1#3$}\hidewidth}}
\egroup
}{%
\bgroup
\scalebox{#4}{\ooalign{$\scriptstyle#2\mkern#1$\cr
\hidewidth{$\scriptstyle\mkern#1#3$}\hidewidth}}
\egroup
}{%
\bgroup
\scalebox{#4}{\ooalign{$\scriptscriptstyle#2\mkern#1$\cr
\hidewidth{$\scriptscriptstyle\mkern#1#3$}\hidewidth}}
\egroup
}
\endgroup
}

\makeatletter
\DeclareRobustCommand{\qto}{%
\begingroup
\m@th
\def\dg@quantumwave{\mathchoice
{\raisebox{.9ex}{$\displaystyle\sim$}}
{\raisebox{.9ex}{$\textstyle\sim$}}
{\raisebox{.9\fontdimen5\scriptfont2}{$\scriptstyle\sim$}}
{\raisebox{.9\fontdimen5\scriptscriptfont2}{$\scriptscriptstyle\sim$}}}%
\mathrel{\vphantom{\dg@quantumwave}\overlay{\rightarrow}
{\mkern-2.25mu\dg@quantumwave\mkern2.25mu}}%
\endgroup
}
\makeatother
\NewcommandTwoArgsLastOpt{\GF}
{{\mathcal{G\!F}_{#1}}}
{{\mathcal{G\!F}_{#1}^{#2}}}

\def\?{\mskip 0.5\thinmuskip}%

\makeatletter
\ifthenelse{\boolean{are_slides}}{

\NewcommandTwoArgsLastOpt{\ssframe}
{\only<1>{\ssect{#1}}\frametitle{#1}}
{\only<1>{\ssect{#1}[#2]}\frametitle{#1}}

\newdimen\progressbar@width
\newcount\progressbar@mypage
\newcount\progressbar@topage

\setbeamertemplate{footline}{
\color{dgProbar}

}

\setbeamercolor{background canvas}{bg=dgPreBack}
}{}%
\makeatother

\numberwithin{theorem}{section}

\renewenvironment{abstract}{%
\small
\begin{center}
{\bfseries\abstractname\vspace{-.5em}\vspace{0pt}}
\end{center}
\list{}{\leftmargin1em\rightmargin\leftmargin}
\item\relax
}{\endlist}

\makeatletter
\let\dg@articlemaketitle\maketitle
\renewcommand{\maketitle}{{\let\rlap\hbox\dg@articlemaketitle}}
\makeatother

\hypersetup{
pdftitle={BQPcc is not contained in MAcc},
pdfauthor={Dmytro Gavinsky}}

\NewM{\CheatShape}{\mathsf{Cheat\text{-}Shape}}
\NewM{\BQPcc}{\mathsf{BQP}^{\mathrm{cc}}}
\NewM{\MAcc}{\mathsf{MA}^{\mathrm{cc}}}

\title{$\mathsf{BQP}^{\mathrm{cc}}$ is not contained in $\mathsf{MA}^{\mathrm{cc}}$}
\author{Dmytro Gavinsky\thanks{In 2022 the author changed his first name, as explained on his Internet page.}%
\thanks{This work was developed in collaboration with {\fontencoding{T1}\fontfamily{lmr}\selectfont\textsc{Anna}} -- Anonymous Neural-Network Assistant -- whom the author sincerely thanks.}\\[0.4ex]
{\small Institute of Mathematics of the Czech Academy of Sciences,
\v{Z}itn\'a 25, Praha~1, Czechia}\\[-0.1ex]
{\small IonQ Quantum, Inc., 4505 Campus Drive,
College Park, MD 20740, USA}}
\date{}

\begin{document}
\maketitle
\thispagestyle{empty}

\abstart
\noindent We exhibit a bipartite language with a poly-logarithmic quantum communication protocol and polynomial Merlin--Arthur communication complexity.
\abend

\sect{Introduction}

The total function introduced in~\cite{CheatShape} separates quantum from randomised communication.
We show that its complement remains hard classically even when the players receive a short classical proof before they interact -- that is,
\m{
\BQPcc\not\subseteq\MAcc.
}

\sect{Preliminaries}

The cost of a communication protocol is the maximum total number of communicated bits and qubits.
For a Boolean function $f$, let $Q_\varepsilon(f)$ be the minimum cost of a quantum protocol with worst-case error at most $\varepsilon$.
In a public-coin Merlin--Arthur protocol, Merlin supplies both players with the same classical proof before the public random coins are sampled, after which the players run a randomised protocol with arbitrarily many rounds.
For every $1$-input some proof must lead to acceptance with probability at least $1-\varepsilon$, whereas on every $0$-input every proof must lead to acceptance with probability at most $\varepsilon$.
The cost of a protocol is the proof length plus the cost of the verifier and $\mathsf{MA}_\varepsilon(f)$ denotes its minimum over the protocols satisfying the correctness requirements.

A bipartite language is a set $\mathcal L\subseteq\bigcup_{N\ge1}(\{0,1\}^N\times\{0,1\}^N)$, and $\mathcal L_N$ is its characteristic function on inputs of length $N$ per party.
We write $\mathcal L\in\BQPcc$ if
\m{
Q_{1/3}(\mathcal L_N)\in O(\log^d N)
}
for some constant $d$ and define $\MAcc$ analogously using $\mathsf{MA}_{1/3}$.

We recall the statements from~\cite{CheatShape} that we require.
Let $N_n$ be the input length per party of $\CheatShape_n$.
By~\cite[Section~6 and Lemma~6.1]{CheatShape},
\m[eq:length-and-quantum]{
n\le N_n\in O(n^3\log^2n),
\qquad
Q_{1/3}(\CheatShape_n)\in O(\log^3N_n).
}
The quantum protocol has two messages, the first quantum and the second classical.
By~\cite[Lemma~7.2]{CheatShape}, for every sufficiently large $n$ there are distributions $\Lambda_{0,n}$ and $\Lambda_{1,n}$, supported on $\CheatShape_n^{-1}(0)$ and $\CheatShape_n^{-1}(1)$, respectively, and quantities
\m{
A_n\in n^{O(1)},\qquad
\eta_n\in2^{-\Omega(\sqrt n)},
}
such that every rectangle $R$ in the input space satisfies
\m[eq:rectangle-bound]{
\Lambda_{0,n}(R)\le A_n\Lambda_{1,n}(R)+\eta_n.
}
Here a rectangle is a set $\mathcal A\times\mathcal B$, where $\mathcal A$ and $\mathcal B$ are sets of Alice's and Bob's inputs, respectively.

\sect{The Merlin--Arthur lower bound}

\theo[thm:main]
{
The family $G_n=1-\CheatShape_n$ satisfies
\m{
Q_{1/3}(G_n)\in O(\log^3N_n),
\qquad
\mathsf{MA}_{1/3}(G_n)
\in\Omega\!\left(\frac{N_n^{1/12}}{\log^{1/6}N_n}\right).
}
}

The lower-bound argument below is based upon Klauck~\cite{Klauck2003}.

\prfstart[of \theoref{thm:main}]
Consider a Merlin--Arthur protocol for $G_n$ with error at most $1/3$, proof length $a$, and verifier cost $q$, and put $c=a+q$.
Amplify the verifier by independent repetitions using the same proof, reducing its completeness and soundness error to
\m{
\varepsilon=\frac{2^{-a}}{8A_n},
}
assuming without loss of generality that $A_n\ge 2$.
The amplified verifier has cost
\m[eq:amplified-cost]{
q'\in O\bigl(q(a+\log A_n)+1\bigr)
}
and its transcript is assumed to fix the protocol answer.

Choose a witnessing proof for every $1$-input of $G_n$.
Since $\Lambda_{0,n}$ is supported on those inputs and there are fewer than $2^{a+1}$ proofs, some proof $z$ witnesses a set of $\Lambda_{0,n}$-mass at least $2^{-a-1}$.
Fix that proof.
The amplified verifier then accepts with probability at least $(1-\varepsilon)2^{-a-1}$ under $\Lambda_{0,n}$, while soundness bounds its acceptance probability under $\Lambda_{1,n}$ by $\varepsilon$.
These probabilities are over both the input distribution and the verifier's public coins.

For each fixed choice of the coins, the accepting transcripts form at most $2^{q'}$ pairwise disjoint rectangles.
Sum~\eqref{eq:rectangle-bound} over them and then average over the coins to obtain
\m{
(1-\varepsilon)2^{-a-1}
\le A_n\varepsilon+2^{q'}\eta_n.
}
Since $\varepsilon\le1/8$ and $A_n\varepsilon=2^{-a}/8$, this gives
\m{
\frac14\,2^{-a}\le2^{q'}\eta_n,
\qquad
a+q'\in\Omega(\sqrt n).
}
Together with~\eqref{eq:amplified-cost}, $A_n\in n^{O(1)}$, and $c=a+q$, it follows that
\m{
c(c+\log n)\in\Omega(\sqrt n),
\qquad
c\in\Omega(n^{1/4}).
}
The length bound in~\eqref{eq:length-and-quantum} implies the result.
\prfend[\theoref{thm:main}]

\crl
{
$\BQPcc\not\subseteq\MAcc$, as witnessed by $\mathcal L= \bigcup_n G_n^{-1}(1)$ -- assuming without loss of generality that $(N_n)_n$ is strictly monotone.
}

\bibstart
\small
\setlength{\itemsep}{0pt}

\bibitem{CheatShape}
Dmytro Gavinsky.
\newblock On the quantum communication complexity of total functions.
\newblock Full version of \href{https://arxiv.org/abs/2608.18784}{arXiv:2608.18784}, 2026.

\bibitem{Klauck2003}
Hartmut Klauck.
\newblock Rectangle size bounds and threshold covers in communication complexity.
\newblock In \emph{Proceedings of the 18th IEEE Conference on Computational Complexity (CCC)}, pages 118--134, 2003.
\newblock \href{https://doi.org/10.1109/CCC.2003.1214415}{doi:10.1109/CCC.2003.1214415}; extended version \href{https://arxiv.org/abs/cs/0208006}{arXiv:cs/0208006}.

\bibend

\end{document}